\documentclass[10pt,conference]{IEEEtran}
\makeatletter
\def\ps@headings{%
\def\@oddhead{\mbox{}\scriptsize\rightmark \hfil \thepage}%
\def\@evenhead{\scriptsize\thepage \hfil \leftmark\mbox{}}%
\def\@oddfoot{}%
\def\@evenfoot{}}
\makeatother
\usepackage{amssymb, amsmath}
\usepackage[dvipdfmx]{graphicx}
\usepackage{mediabb}
\usepackage{cite}
\usepackage{bm}
\usepackage{ascmac}
\usepackage{cases}
\usepackage{flushend}
\usepackage{wrapfig}
\usepackage[hang,small,bf]{caption}
\usepackage[absolute,showboxes]{textpos}

\usepackage{xcolor}

\def\bq{\begin{equation}}
\def\eq{\end{equation}}
\def\bqn{\begin{eqnarray}}
\def\eqn{\end{eqnarray}}
\def\bqnn{\begin{eqnarray*}}
\def\eqnn{\end{eqnarray*}}

\def\argmax{{\rm argmax}}

\TPMargin{5pt}

\newcommand{\copyrightstatement}{
    \begin{textblock}{0.84}(0.08,0.93)    
         \noindent
         \footnotesize
         \copyright~2020 IEEE.  Personal use of this material is permitted.  Permission from IEEE must be obtained for all other uses, in any current or future media, including reprinting/republishing this material for advertising or promotional purposes, creating new collective works, for resale or redistribution to servers or lists, or reuse of any copyrighted component of this work in other works.
    \end{textblock}
}

\title{Recovery command generation towards automatic recovery in ICT systems by Seq2Seq learning}
\author{
\IEEEauthorblockN{Hiroki Ikeuchi$^{\ast}$, Akio Watanabe$^{\dagger}$, Tsutomu Hirao$^{\ddagger}$, Makoto Morishita$^{\ddagger}$,\\ Masaaki Nishino$^{\ddagger}$, Yoichi Matsuo$^{\ast}$, Keishiro Watanabe$^{\ast}$}
\IEEEauthorblockA{$^{\ast}$NTT Network Technology Laboratories, NTT Corporation, Tokyo 180-8585, Japan}
\IEEEauthorblockA{$^{\dagger}$NTT East Corporation, Tokyo 163-8019, Japan}
\IEEEauthorblockA{$^{\ddagger}$NTT Communication Science Laboratories, NTT Corporation, Kyoto 619-0237, Japan}
Email: $^{\ast}$\{hiroki.ikeuchi.re, yoichi.matsuo.ex, keishiro.watanabe.ry\}@hco.ntt.co.jp,\\
$^{\ddagger}$\{tsutomu.hirao.kp, makoto.morishita.gr, masaaki.nishino.uh\}@hco.ntt.co.jp,
 $^{\dagger}$akio.watanabe.ht@east.ntt.co.jp
}

\date{}

\begin{document}
\copyrightstatement

\maketitle
\begin{abstract}
With the increase in scale and complexity of ICT systems, their operation increasingly requires automatic recovery from failures.
Although it has become possible to automatically detect anomalies and analyze root causes of failures with current methods,
making decisions on what commands should be executed to recover from failures still depends on manual operation, which is quite time-consuming. Toward automatic recovery, we propose a method of estimating recovery commands by using Seq2Seq, a neural network model.
This model learns complex relationships between logs obtained from equipment and recovery commands that operators executed in the past.
When a new failure occurs, our method estimates plausible commands that recover from the failure on the basis of collected logs.
We conducted experiments using a synthetic dataset and realistic OpenStack dataset, demonstrating that our method can estimate recovery commands with high accuracy.

\end{abstract}


\section{Introduction}
Automatic recovery from failures has been necessary in recent information and communications technology (ICT) systems.
As ICT systems become larger and more complex,
a wider variety of failures will occur and their number will increase.
Under such circumstances, manual operations are time-consuming and have been unable to keep system downtime short.

A great deal of effort has been made to automate failure handling.
Log management platforms, such as \cite{splunk,Elasticsearch}, automatically collect and analyze {\it logs}, 
i.e., text data to inform operators of various symptoms of system status, such as syslogs or alarms (e.g., Figure~\ref{log-command-example} (upper)).
On the basis of these logs, various methods have been proposed to automate the processes of anomaly detection~\cite{anomaly_detection} and root-cause analysis~\cite{ikeuchi}.
However, the {\it recovery process} is still done manually, which involves two steps; determining which commands ({\it recovery commands}) should be executed to recover from a failure and executing the determined recovery commands (e.g., Figure~\ref{log-command-example} (lower)).
The latter step can be automated using run book automation (RBA) tools, but the former needs to be done by operators and is quite time-consuming.
Although there are studies~\cite{ticket1,ticket2,ticket3,workflow1,workflow2,workflow3} that suggest recovery actions for operators,
most rely on trouble tickets, which do not always include sufficient information to precisely determine recovery commands.
As such, we still cannot free operators from the task of recovering from failures in spite of much research on the topic.

Towards automatic recovery, we are aiming toward developing a method of automatically generating the recovery commands directly from logs.
If the automatic generation of recovery commands is successful, the whole recovery process can be accomplished automatically just by executing the commands with RBA tools.
Since logs and recovery commands usually accumulate in the equipment, we take an approach in which we use such accumulated data.
\begin{figure}[b]
\begin{center}
\begingroup
\renewcommand{\arraystretch}{0.7}
\begin{tabular}{|l||c|}\hline
\multicolumn{1}{|c||}{\fontsize{8pt}{0cm}\selectfont log message} & \multicolumn{1}{|c|}{\fontsize{8pt}{0cm}\selectfont ID}\\ \hline\hline
{\fontsize{6pt}{0cm}\selectfont 14:00:00 proc01 DEBUG [req-12345] accepted ( IPv4, 12345) server /***/***/***}&{\fontsize{6pt}{0cm}\selectfont 1}\\
{\fontsize{6pt}{0cm}\selectfont 14:00:01 proc02 INFO [req-56789] Get http://***}&{\fontsize{6pt}{0cm}\selectfont 2}\\
{\fontsize{6pt}{0cm}\selectfont 14:00:03 proc01 DEBUG [req-24680] Failed to fetch instance by id server1 get /***/***}&{\fontsize{6pt}{0cm}\selectfont 4}\\
{\fontsize{6pt}{0cm}\selectfont 14:00:03 proc01 DEBUG [req-13579] Returning 404 to user: Could not find instance ***}&{\fontsize{6pt}{0cm}\selectfont 5}\\
{\fontsize{6pt}{0cm}\selectfont 14:00:03 proc01 DEBUG [req-98765] HTTP exception thrown: Could not find instance ***}&{\fontsize{6pt}{0cm}\selectfont 7}\\
{\fontsize{6pt}{0cm}\selectfont 14:00:04 proc01 DEBUG [req-43210] Returning 404 to user: Could not find instance ***}&{\fontsize{6pt}{0cm}\selectfont 5}\\ \hline
\end{tabular}
\endgroup
\\
\vspace*{0.3cm}
\begingroup
\renewcommand{\arraystretch}{1}
\begin{tabular}{|l|}\hline
{\fontsize{10pt}{0cm}\selectfont openstack-status \verb+|+ grep down}\\
{\fontsize{10pt}{0cm}\selectfont systemctl restart nova-scheduler}\\
{\fontsize{10pt}{0cm}\selectfont openstack-status \verb+|+ grep scheduler}\\ \hline
\end{tabular}
\endgroup
\end{center}
\caption{(Upper) Example of log messages and assigned IDs of OpenStack system. Since request IDs are uninformative variables, log messages with same template and different request IDs (req-13579 and req-43210) have same log ID 5. (Lower) Recovery commands of OpenStack system with nova-scheduler failure.}
\label{log-command-example}
\end{figure}

To generate recovery commands using past data, the following challenges need to be addressed.
First, logs and recovery commands are non-deterministic and not in one-to-one correspondence.
Even if the same failure occurs as before, logs may fluctuate and the recovery commands may differ depending on the operator.
Second, the length of recovery commands might differ depending on the failure.
This makes it difficult to apply simple classification techniques to this problem.
Third, the reliability of estimation needs to be measured.
This is necessary from a practical point of view because operators need to determine whether it is safe to execute the estimated commands automatically on the basis of reliability.

Taking such challenges into account, we propose a method of estimating recovery commands by using sequence-to-sequence (Seq2Seq)~\cite{Seq2Seq,attention},
a neural network model usually used to solve translation tasks in the field of natural language processing.
In translation tasks, Seq2Seq can learn the relationship between sentences in different languages and convert an input sentence in one language into that in another language.
Although the correspondence of a sentence to its translated sentence is not always one-to-one and the lengths of sentences are variable, Seq2Seq can successfully acquire a translation model with translation databases.
This is very similar to our problem setting and that is why we expect that Seq2Seq has the capability of solving the first and second challenges described above.
Using accumulated data as training data, our method trains Seq2Seq representing the complex relationship between two non-deterministic length-variable sequences, i.e., logs and recovery commands.
When a new failure occurs during operation, the trained Seq2Seq infers plausible recovery commands on the basis of the observed logs.
Regarding the third challenge, we can adopt the likelihood of the estimated commands as the reliability.
We verified the effectiveness of our method by conducting experiments using synthetic and realistic datasets obtained from an OpenStack~\cite{openstack} system, demonstrating that our method can estimate recovery commands with high accuracy.

The rest of this paper is organized as follows. In Section~\ref{method}, we state the problem of estimating recovery commands and introduce the proposed method based on Seq2Seq. In Section~\ref{experiments}, we discuss the effectiveness of our method validated through experiments using a synthetic dataset and a realistic OpenStack dataset. In Section~\ref{related_work}, we introduce related work. We conclude this paper and mention future work in Section~\ref{conclusion}.

\section{Problem statement and proposed method}\label{method}

%


We describe a problem of recovery command estimation.
We assume a training dataset consisting of $N$ pairs of logs and recovery commands $D=\{({logs}^{(k)},cmds^{(k)})\}_{k=1}^{N}$.
Now, we consider a situation in which a new failure occurs and new logs ${logs}^{(test)}$ are obtained. Then, the recovery command estimation involves estimating appropriate recovery commands by using a certain estimator $f$ such as $f({logs}^{(test)})={cmds}^{(test)}$. We regard the estimation as successful if the system is recovered by executing ${cmds}^{(test)}$.

In Section~\ref{subsec-Seq2Seq}, we briefly review Seq2Seq, and in Section~\ref{subsec-proposed}, we explain how to apply it to our problem.

\subsection{Seq2Seq}\label{subsec-Seq2Seq}
Seq2Seq~\cite{Seq2Seq, attention} is a neural network model that learns the relationship between input and output sequences,
and has been widely used as a basic model of translation systems. 
Seq2Seq converts a source sequence ${\bf X}=\langle X_1, X_2, \dots X_{|{\bf X}|}\rangle$ into a target sequence ${\bf Y}=\langle Y_1, Y_2, \dots Y_{|{\bf Y}|}\rangle$ in the following manner.

First, ${\bf X}$ is converted into a single hidden vector $c$ by an encoder.
More precisely, after each word of ${\bf X}$ is embedded in a fixed dimensional space, all are recurrently mapped into $c$ by using a recurrent neural network (RNN) such as $\bar{h}_t=RNN_{enc}(X_s, h_{s-1}), \bar{h}_0=\bm{0}, c=\bar{h}_{|{\bf X}|}$.
Then, $c$ is recurrently unfolded and generates one target word $Y_t$ at a time by using a decoder.
Formally, we express the decoder as follows:
\bqnn
h_t=RNN_{dec}(Y_{t-1}, h_{t-1}),\quad&\mbox{(decoder RNN)}\\
\tilde{c}_t = ATTN(h_t, \{\bar{h}_{s}\}_{s=1}^{|{\bf X}|}),\quad&\mbox{(decoder attention)}\\
p(Y|Y_{<t}, {\bf X})=OUT(\tilde{c}_t, h_t),\quad&\mbox{(decoder output)}\\
Y_t=\argmax_{Y}p(Y|Y_{<t}, {\bf X}),
\eqnn
where $s_0=c$, $Y_{<t}=\langle Y_1,Y_2,\dots Y_{t-1}\rangle$, and $Y_0$ and $Y_{|{\bf Y}|}$ are special symbols indicating the start and end of a sentence, respectively. $ATTN(\cdot)$ indicates an attention layer that computes the context vector $\tilde{c}_t$, which is designed to refer to all the source hidden states $\{\bar{h}_s\}_s$ and represents how much ``attention'' should be paid to each $\{\bar{h}_s\}_s$ to estimate $Y_t$. $OUT(\cdot)$ finally outputs the conditional probability on $Y_t$. 
We refer the reader to the study by Luong et al.~\cite{attention} for more details.
In the learning phase, Seq2Seq acquires a model that maximizes the (log) likelihood of an output sequence, i.e., $\log\Pi_t p(Y_t|Y_{<t}, {\bf X})$.

\begin{figure}[tbp]
  \begin{center}
\includegraphics[width=80mm]{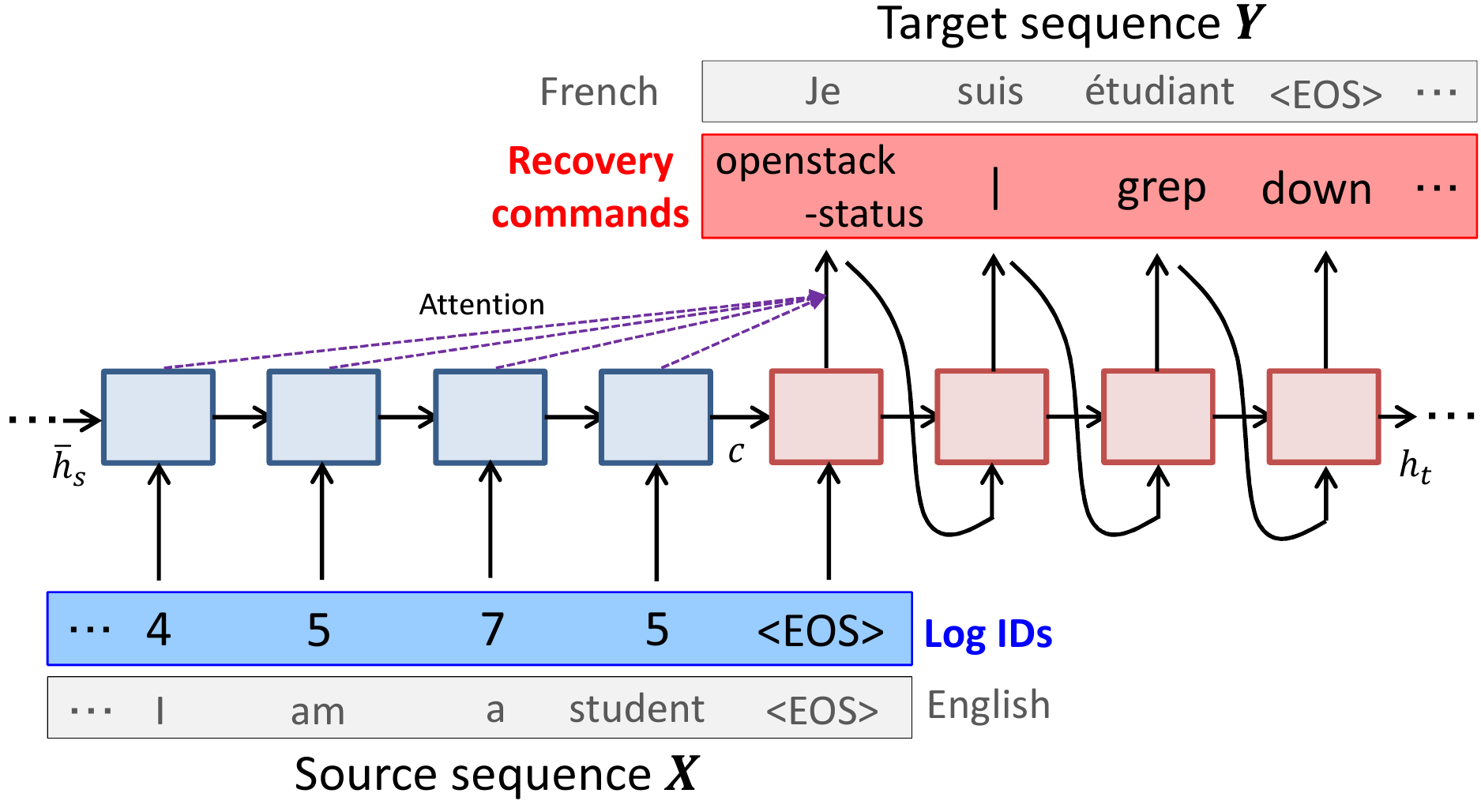}
  \end{center}
\caption{Recovery command estimation (our method) and English-French translation (conventional usage) by using Seq2Seq}
  \label{fig_seq2seq}
\end{figure}

\subsection{Proposed method}\label{subsec-proposed}
The main idea of our method is to use Seq2Seq as $f$. Figure~\ref{fig_seq2seq} shows an overview of our method compared with English-French translation, a conventional use of Seq2Seq.
To use Seq2Seq as $f$, we need to render logs and recovery commands in the sequential form.
Since the recovery commands consist of a relatively small number of words, we can simply consider them as a sequence of words ${\bf Y}=\langle Y_1, Y_2, \dots Y_{|{\bf Y}|}\rangle, Y_i\in \bm{V}$, where $\bm{V}$ is a set of words appearing in all commands.
The amount of words included in logs, however, is too large to directly input to Seq2Seq.
The longest sentence handled by Sutskever et al.~\cite{Seq2Seq} consists of 79 words, while the number of words in logs to be dealt with usually exceeds 5,000 (when we assume that one log message includes 50 words and more than 100 log messages are generated from one device within one minute).
In addition, logs include uninformative variable parameters to determine recovery commands, such as date, time or detailed process ID, which may disrupt the training and estimation. 
To overcome these difficulties, we use a log-template-extraction tool~\cite{kimura}, with which we mask the uninformative variable parameters in log messages, extract ``log templates'', which are the remaining invariable parts of log messages, and assign an ID number to each log template. Thanks to this abstraction, we can reduce unnecessary dimensions of input data and treat logs as a sequence of natural numbers; ${\bf X}=\langle X_1, X_2, \dots X_{|{\bf X}|}\rangle, X_i\in\mathbb{N}$.
In the example of Figure~\ref{log-command-example}, we have ${\bf X}=\langle 1,2,4,5,7,5\rangle$，${\bf Y}=\langle$openstack-status, \verb+|+, grep, down, \verb|<|ENT\verb|>|, systemctl, restart, nova-scheduler, \verb|<|ENT\verb|>|, openstack-status, \verb+|+, grep, scheduler, \verb|<|EOC\verb|>|$\rangle$. Here, \verb|<|ENT\verb|>| is a word indicating the pressing of the enter key and \verb|<|EOC\verb|>|\ the end of the series of commands. Note that log messages with the same template and different invariable parameters have the same ID 5 assigned.

To determine whether to execute the estimated recovery commands, we also need to quantify the reliability of the estimation.
We can consider the likelihood $\Pi_t p(Y_t|Y_{<t}, {\bf X})$ as the probability that the commands ${\bf Y}$ would be executed given logs ${\bf X}$. Thus, we adopt $\left(\Pi_t p(Y_t|Y_{<t}, {\bf X})\right)^{1/|{\bf Y}|}$ as the reliability of the estimated recovery commands.

\section{Experiments}\label{experiments}
We validated our method using two types of datasets, a synthetic dataset and realistic OpenStack dataset.

\subsection{Configuration}
We implemented Seq2Seq using the translation model in fairseq~\cite{fairseq}.
Instead of RNN cells, we used long short-time memory (LSTM) cells.
LSTM is an extension of RNN that enables the learning of long-term dependencies.
We set the dimension of the input/output embedding layer to 512, that of  the hidden vector to 512, and the dropout probability to 0.2.
We continued to train the models until no improvement in loss for validation was observed in the last ten epochs.
In the inferring phase, the decoding beam size was set to 5.

\subsection{Synthetic dataset}\label{synthetic_dataset}
We prepared a synthetic dataset by imitating realistic logs and recovery commands.
The main purposes of this experiment were to see whether our method can handle various types of recovery commands, such as rollback commands or typos, and whether it is robust against randomness of logs and recovery commands.

In consideration of a diversity of real failures, we considered a total of 50 types of failures and divided them into 5 groups (A)$\sim$(E), each of which includes 10 types of failures.
We assume that failures that belong to the same group have similar properties and their generated logs and recovery commands are also similar.

We now explain how to construct the dataset of logs and recovery commands in detail.\\
{\bf Logs:}
In creating logs, we considered the following properties of log appearance. 
(i) Similar failures generate a large number of the same logs and a few different logs.
For example, consider two failures that have the same root cause (e.g., termination of a particular process) occurring in different servers.
They generate almost the same logs because the root cause is common but also generate a few distinct logs that include each host name.
(ii) Logs unrelated to failures randomly get mixed in with logs of failures. They are generated by cron jobs, periodic normality check, or for some other (usually unknown) reasons. (iii) The order of log appearance is subject to many factors, such as system response or network delay, and may randomly change.

\begin{figure}[htbp]
\centerline{\includegraphics[scale=0.4,angle=270]{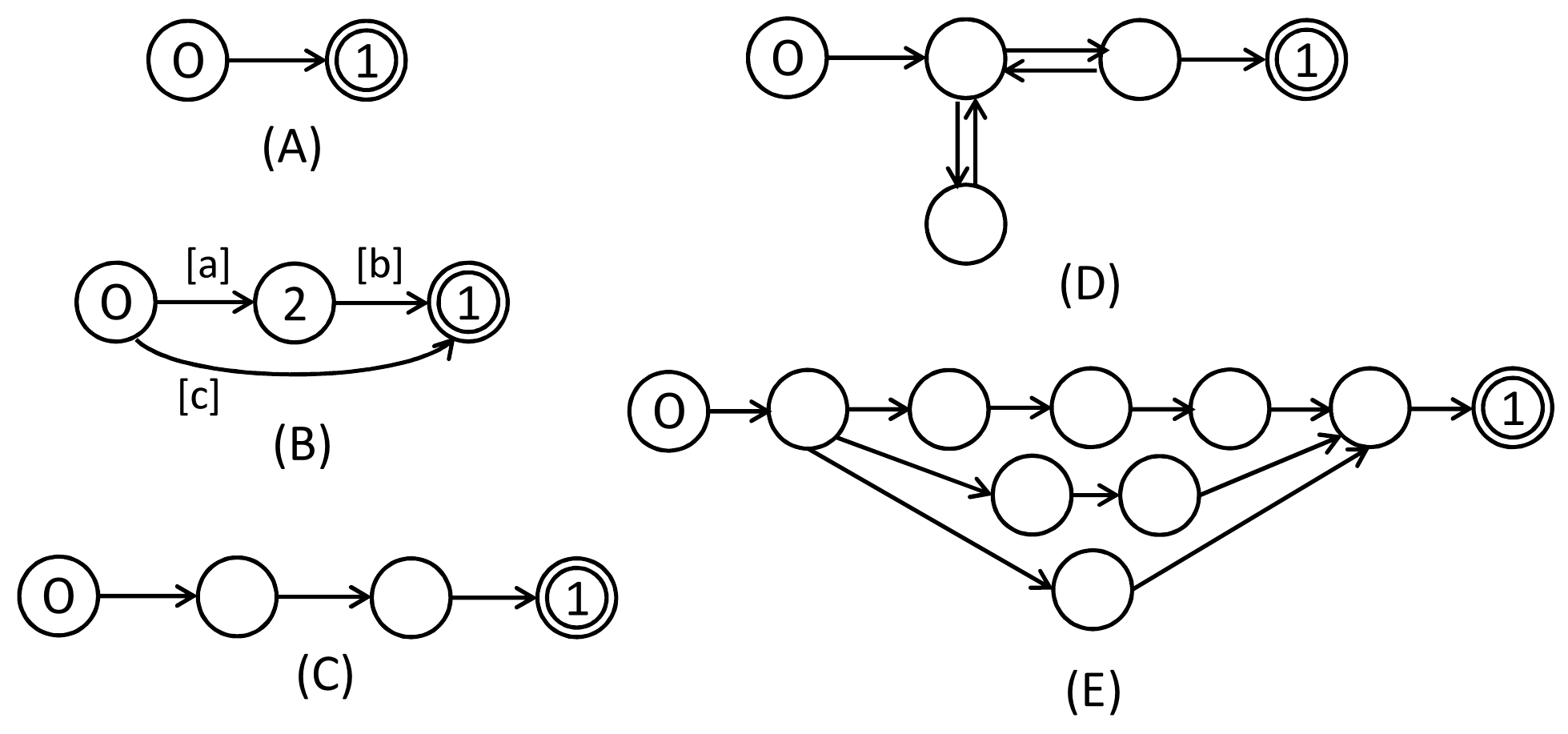}}
\caption{Automata that represent recovery measures of groups (A)$\sim$(E). For example, transition [a] in automaton (B) occurs when
current state is 0 (initial failed state) and one of three correct commands listed in Table~\ref{ex_cmd} is executed.}
\label{automata}
\end{figure}

Now, we explain how to create logs (sequences of log IDs) of group (A). We did the same for the other groups (B)$\sim$(E).
First, we randomly generated a sequence of natural numbers the length of which is about 150.
In accordance with property (i), we then created ten different sequences corresponding to ten types of failures in group (A) by adding a few distinct numbers in random positions of the original sequence. For each sequence, we also added noisy numbers in random positions and randomly exchanged the orders of adjacent numbers in consideration of properties (ii) and (iii). As a result, each sequence consisted of about 200 IDs.\\
{\bf Recovery commands:}
In general, if we execute a certain command, we can consider that the state of the ICT system changes in some way.
Thus, the process of executing recovery commands line by line can be considered a state transition process that starts from a ``failure state'' and reaches a ``recovered state''. To express various types of recovery measures, we constructed five types of state transition processes, i.e., automata.
Figure~\ref{automata} shows five automata (A)$\sim$(E) corresponding to groups (A)$\sim$(E).
In each automaton, state 0 is the initial failed state and state 1 is the recovered state (accept state). Each arrow indicates a transition that occurs if and only if a {\it correct command} is executed. We assigned each arrow several types of correct commands because recovery commands are not determined uniquely and may differ depending on the operator. If an incorrect command is executed, the current state does not change.
As an example, Table~\ref{ex_cmd} shows the correct commands of each transition in automaton (B). The transitions [a], [b], and [c] correspond to the arrows with the same labels in Figure~\ref{automata}. \verb|<|FailedComponent\verb|>|$\in$\{cmp1, cmp2,...,cmp10\} is a placeholder that has a different word depending on the failure in group (B).
If the failure belongs to group (B) and the failed component is cmp1, the recovery commands $\langle$cmd1, xxx, cmp1, \verb|<|ENT\verb|>|, show, status \verb|<|ENT\verb|>|, cmd2, restart, cmp1, \verb|<|EOC\verb|>|$\rangle$ can successfully recover from the failure, while $\langle$cmd1, xxx, cmp1, \verb|<|ENT\verb|>|, cmd1, start, cmp1, \verb|<|EOC\verb|>|$\rangle$ cannot because the system is still in state 2.

Each type of automata is designed to represent realistic recovery measures.
Automata (A)$\sim$(C) are relatively simple, and in reality, most recovery measures can be expressed like these. The recovery commands in Section~\ref{realistic} can be considered as (A)$\sim$(C). Automaton (D) has arrows of opposite directions, which indicates rollback actions.
Automaton (E) has a branch and several routes to recover from a failure. 
Experts commonly take the shortest path to recover, while beginners take a route requiring much time and more actions.

By using these automata, we created recovery commands as follows.
First, by randomly choosing any one of the correct commands in each transition and connecting them, we created an arbitrary sequence of correct commands that is accepted by a given automaton.
Then, we randomly inserted several incorrect commands in the accepted command sequence.
The inserted incorrect commands represent typos or commands for status check such as ``show status''.
We repeated the above procedure for every failure of every group.

To investigate the dependence of our method on the amount of data, we trained Seq2Seq while varying the number of samples included in the training and development set (90:10 split).
After training, we evaluated the success rate, which is defined as the ratio of the number of recovery commands accepted by the corresponding automaton to 450, the total number of samples in the test set.

\begin{table}[h]
	\caption{Correct commands in automaton (B)}
    \begin{center}
    \begin{tabular}{|c||l|} \hline
		Transition & Correct command\\ \hline
	    [a] in (B) & cmd1 -a xxx \verb|<|FailedComponent\verb|>| \\ 
	     & cmd1 -b xxx \verb|<|FailedComponent\verb|>| \\ 
	     & cmd1 xxx \verb|<|FailedComponent\verb|>| \\ \hline
	    [b] in (B) & cmd2 start \verb|<|FailedComponent\verb|>| \\ 
		& cmd2 restart \verb|<|FailedComponent\verb|>| \\ \hline
	    [c] in (B) & reboot \\ 
		& shutdown -r now \\ \hline
    \end{tabular}
    \end{center}
    \label{ex_cmd}
\end{table}

\subsubsection*{[Results]}
Table~\ref{results1} shows the success rates averaged over five trials.
The results are averaged over ten types of failures in each group.
When we have a sufficiently large amount of training data, we found that our method can almost correctly estimate recovery commands for every type of failure.
We also have two significant findings by looking into the estimated commands.
One is that the estimated recovery commands did not include any incorrect commands such as typos.
This is a significant advantage of our method.
The other is that our method did not always output the shortest paths (the minimum recovery commands).
For example, some estimated commands in group (D) included rollback commands which do not appear in the shortest paths.
If we want to avoid such recovery commands, it may be a good idea to estimate several patterns of recovery commands using beam search and choose the shortest ones.

As the amount of training data decreased, the success rates started to drop; suddenly for groups (C), (D), and (E).
Since the path lengths from state 0 to state 1 for automata (C), (D), and (E) are longer, there is a wider range of variations in recovery commands.
This is why groups (C), (D), and (E) require more training data to correctly estimate recovery commands.
In practice, if we need to deal with such failures, it may be a good idea to increase the amount of data by fault injection in the verification system.

%

\begin{table}[h]
	\caption{Successful rates of estimated recovery commands. Results are averaged over ten types of failures in each group.}
	\label{accuracy_rate}
    \begin{center}
    \begin{tabular}{|c||c|c|c|c|c|} \hline
	\shortstack{\# samples of train-dev\\ (per failure)} &
	\shortstack{Group \\ (A)} &
	\shortstack{Group \\ (B)} &
	\shortstack{Group \\ (C)} &
	\shortstack{Group \\ (D)} &
	\shortstack{Group \\ (E)} \\ \hline
	4500 (90) samples& 91.6\% & 91.7\% & 71.1\% & 72.1\% & 75.1\% \\ \hline
	3750 (75) samples& 66.4\% & 75.3\% & 3.54\% & 13.1\% & 0.66\% \\ \hline
	3000 (60) samples& 45.8\% & 45.1\% & 4.44\% & 10.7\% & 2.22\% \\ \hline
    \end{tabular}
    \end{center}
    \label{results1}
\end{table}

\vspace*{-0.6cm}
\subsection{Realistic dataset}\label{realistic}
We constructed an OpenStack system for validation.
OpenStack is a good example for applying our method because OpenStack consists of dozens of components and the recovery command determination based on a large volume of logs generated from these components is a tremendously exhausting task.
We considered the 13 failures listed in Table~\ref{list}, which imitate ``component unavailable'' failures accounting for about 52\% of all failures in OpenStack~\cite{logan}.
We manually injected one of the 13 failures and collected logs, the lengths of which were around 100$\sim$600, generated in two minutes.
We repeated this procedure several times for each failure.
Since some user actions, such as virtual machine creation, were sometimes executed within the two minutes, the generated logs differed
depending on the sample even for the same failure.
We also prepared several types of recovery commands for each failure.
This reflects the fact that the recovery commands may differ depending on the operator even for the same failure.
As an example, Figure~\ref{log-command-example} shows some of the logs and recovery commands for the failure ``{\texttt{nova-scheduler}} stop''.

We trained Seq2Seq with training data consisting of 1170 pairs and development data consisting of 130 pairs.
We determined whether the estimation was successful for 121 test cases and investigated the dependence of success rate on reliability.

\begin{table}[h]
\caption{List of 13 failures considered in experiment using realistic dataset}
\label{state}
	\begin{center}
	\begin{tabular}{|l|l|} \hline
	{\bf\texttt{nova-api}} stop & {\bf\texttt{cinder-volume}} stop \\
	{\bf\texttt{nova-scheduler}} stop & {\bf\texttt{cinder-scheduler}} stop \\
	{\bf\texttt{nova-conductor}} stop & {\bf\texttt{neutron-server}} stop \\
	{\bf\texttt{nova-compute}} stop & {\bf\texttt{neutron-openvswitch}} stop \\
	{\bf\texttt{glance-api}} stop & {\bf\texttt{neutron-dhcp-agent}} stop \\
	{\bf\texttt{glance-registry}} stop & {\bf\texttt{neutron-l3-agent}} stop \\
	{\bf\texttt{cinder-api}} stop & \\	\hline
	\end{tabular}
	\end{center}
	\label{list}
\end{table}


\subsubsection*{[Results]}
\begin{figure}[htbp]
\centerline{\includegraphics[scale=0.4]{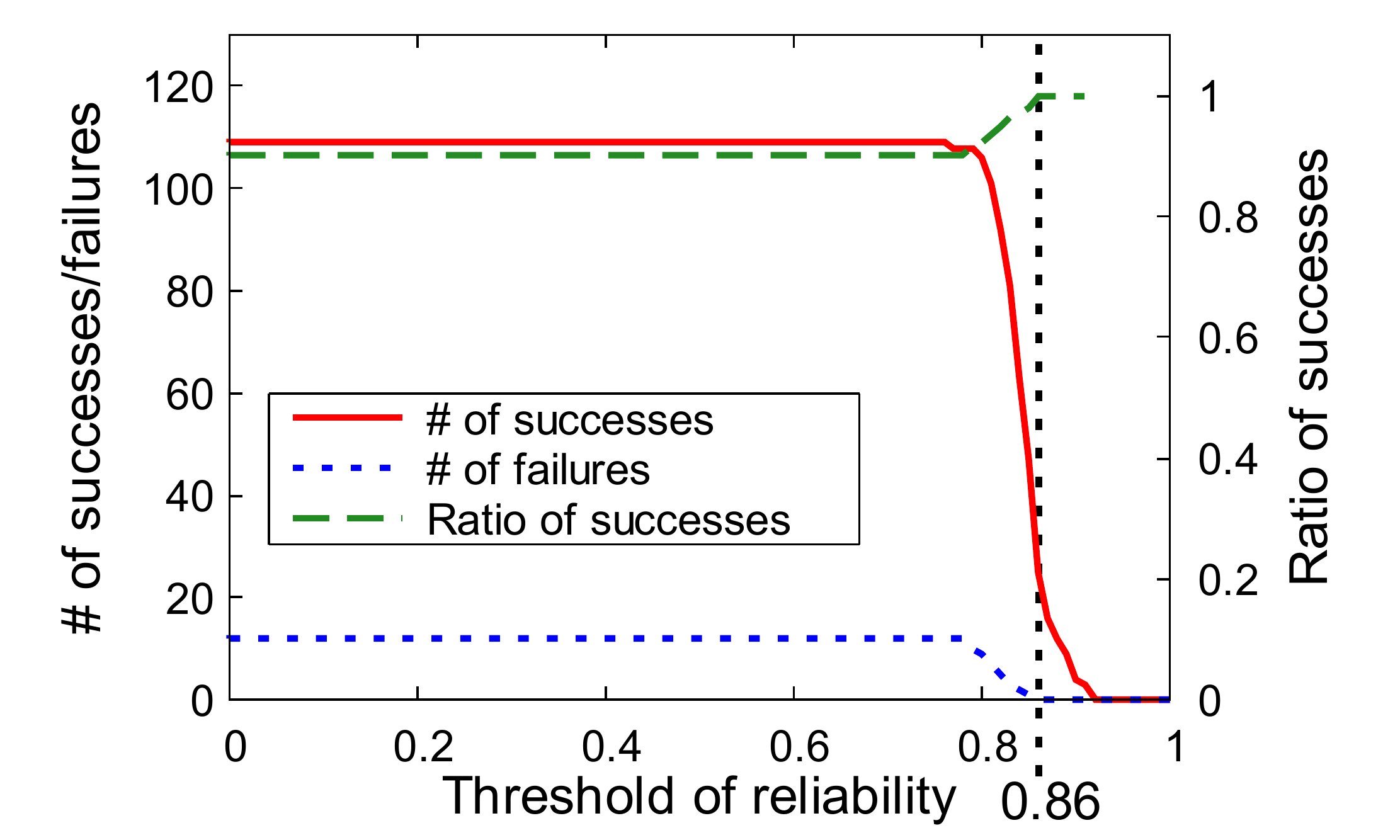}}
\caption{Number of successes/failures (red solid line/dotted blue line) and ratio of successes (broken green line) under given reliability.
Under condition that threshold of reliability is above 0.86 (vertical black dotted line), ratio of successes becomes one, which means all estimated recovery commands satisfying this condition could be executed safely.}
\label{graph}
\end{figure}

Figure~\ref{graph} plots the number of successful/unsuccessful recovery commands under the condition that their reliability is more than a given value (red solid line/dotted blue line, respectively).
Note that the value on the vertical axis (the value at threshold of reliability $=0$) indicates the total number of (un)successful recovery commands.
From this figure and the results of estimated commands, we have the following findings. First, aside from reliability, the success rate was quite high ($109/121\simeq90.1\%$), which demonstrates the effectiveness of our method. Second, the number of (un)successful cases remained almost constant when the threshold of reliability ran from 0 to 0.75. This implies that almost all the estimated commands had a reliability of over 0.75 regardless of success or failure. The definition of reliability may leave room for reconsideration. Third, all the failed estimated commands were syntactically correct as commands, but were recovery commands for different types of failure.

From a practical point of view, it is important for operators to know what the threshold value of reliability should be.
Operators are supposed to determine that they can execute the estimated commands if the reliability is beyond the given threshold.
Operation should generally err on the side of caution because executing inappropriate commands not only fails to recover the system but also may negatively affect the system.
Thus, the ratio of the number of successes to the total number of tests that give reliability over the threshold,
which is denoted with the green broken line in Figure~\ref{graph}, should be one.
As far as this experiment is concerned, we conclude that we should set the threshold of reliability to at least 0.86.


\section{Related work}\label{related_work}

\subsection{Trouble-ticket-based method}
To the best our knowledge, this is the first paper that addresses the estimation of recovery commands without manually predefining recovery actions. Most previous research on failure recovery relied on trouble tickets, which are reports of failure operation written in a natural language. Some methods~\cite{ticket1,ticket2,ticket3} present operators trouble tickets relevant to a failure by using machine learning techniques such as topic models.
Other methods~\cite{workflow1,workflow2,workflow3} analyze multiple trouble tickets (or official troubleshooting guidelines if any) and visualize them as a workflow chart that operators can understand easily. These methods shorten recovery time by supporting operators' decision making.
However, the writing style of trouble tickets greatly differs depending on the operator, and trouble tickets do not always include sufficient information to precisely decide recovery actions.
Thus, an operator needs to read the recommended tickets or workflow charts carefully and determine what action should be taken at his/her discretion.
On the other hand, our method focuses on recovery commands, which do not have such ambiguity and are executable by machines.

\subsection{Optimal recovery action selection}
Intelligent selection of optimal recovery actions is another direction of research on failure recovery.
Some such methods~\cite{pomdp1,pomdp2} formulate the decision-making problem of optimal recovery actions as a partially observed Markov decision model (POMDP). By solving the POMDP problem, they suggest the optimal actions on the basis of the monitored information such as health check of a server or network connection.
They seem compatible with the concept of automatic recovery.
However, they need to construct a probability model imitating the system behavior in advance.
This is rather difficult in the case of complex systems such as OpenStack.
Other promising methods for automatic recovery include reinforcement-learning-based methods.
The methods proposed by Littman et al.~\cite{rl1,rl2} are aimed at acquiring a time-efficient recovery policy by attempting various episodes of actions.
They require a verification system or simulator that imitates a target system.
The method proposed by Zhu and Yuan~\cite{rl3} generates an efficient recovery policy on the basis of the structured reports of recovery history issued by machines.
Although this method does not require a verification system, the structured reports do not always exist unlike logs, which are always accessible as a standard for machine messages. Moreover, all these methods need to predefine the action space, i.e., what commands can be used.

\subsection{Self-healing functions of cloud-based systems}
The relation between our method and self-healing functions with which modern cloud-based systems are equipped is worth mentioning.
In the case of Kubernetes~\cite{k8s}, for example, when a certain failure occurs in physical servers or virtual instances, i.e., pods, the controller manager automatically detects the event and recreates or relocates instances in other servers.
We believe that our method is a complement to such self-healing functions.
In practice, there are failures that cannot be handled with self-healing functions.
For example, problems occurring in lower layers above which a cloud computing service is implemented, such as disconnection of a network through which Kubernetes nodes are connected, need to be resolved with other mechanisms than self-healing functions.
Our method has potential to address such problems as long as accumulated logs and recovery commands are accessible.

\subsection{Seq2Seq learning}
Due to its power and flexibility, Seq2Seq have been applied to a variety of tasks.
A series of studies closely related to ours applied the models of generating source code of general purpose programming languages such as Python~\cite{yin,oda}.
Compared with these tasks, our task is mainly different in the point that its input is highly regularized log data, which means appropriate abstractions would contribute to better performance.

\section{Conclusion}\label{conclusion}
We proposed an automatic recovery command estimation method based on Seq2Seq
and evaluated its effectiveness through experiments using a synthetic dataset and realistic OpenStack dataset.
We also discussed the appropriate setting of the threshold of reliability to automatically execute the estimated commands.
We believe that our method presents a new direction for automatic operation of ICT systems.

Our method still has the following limitations, which remain as for further study. 
(1) We evaluated our method in a small-scale verification system. We need to confirm the effectiveness of our method for larger commercial products by conducting further evaluation. As investigated in Section~\ref{synthetic_dataset}, whether the amount of real data and the fluctuation in real recovery commands are acceptable for Seq2Seq seems to be one of the applicable conditions of our method.
(2) The recovery commands often include variant arguments such as IP addresses or host names. Such information needs to be estimated and filled in appropriately to execute the commands.
Since there are many techniques to estimate such information, we need to develop a method that can collaborate with such techniques.
(3) In contrast to natural language, commands have rigorous syntax and would never work if they broke the syntax even a little. 
In our experiments, we did not observe such wrong commands, but it would be possible in other cases.
Thus, we need to develop a method of incorporating the syntactic information of commands.
(4) Our method infers the recovery commands to the end of recovery at all once.
However, recovery processes sometimes branch depending on the outcome of the executed command.
Thus, we need to develop another method for generating commands iteratively in response to the outcome of the previous commands.

\end{document}